\documentclass[11pt]{revtex4}    
\usepackage{amssymb,epsf}
\usepackage{latexsym}

\begin{document}

\title{Asymptotically flat charged rotating dilaton black holes in higher dimensions}
\author{A. Sheykhi$^{1,2}$\footnote{sheykhi@mail.uk.ac.ir}, M.
Allahverdizadeh$^{1}$, Y.
Bahrampour$^{3}$\footnote{bahram@mail.uk.ac.ir} and M.
Rahnama$^{1}$\footnote{Majid.Rahnama@mail.uk.ac.ir}}
\address{$^1$Department of Physics, Shahid Bahonar University, P.O. Box 76175, Kerman, Iran\\
         $^2$Research Institute for Astronomy and Astrophysics of Maragha (RIAAM), Maragha, Iran\\
         $^3$Department of Mathematics, Shahid Bahonar University, Kerman, Iran}
\begin{abstract}
We find a class of asymptotically flat slowly rotating charged
black hole solutions of Einstein-Maxwell-dilaton theory with
arbitrary dilaton coupling constant in higher dimensions. Our
solution is the correct one generalizing the four-dimensional case
of Horne and Horowitz \cite{Hor1}. In the absence of a dilaton
field, our solution reduces to the higher dimensional slowly
rotating Kerr-Newman black hole solution. The angular momentum and
the gyromagnetic ratio of these rotating dilaton black holes are
computed. It is shown that the dilaton field modifies the
gyromagnetic ratio of the black holes.

\end{abstract}
\maketitle 

\section{Introduction}
General Relativity in higher dimensions has been the subject of
increasing attention in recent years. There are several
motivations for studying higher dimensional Einstein's theory, and
in particular its black hole solutions. First of all, string
theory contains gravity and requires more than four dimensions. In
fact, the first successful statistical counting of black hole
entropy in string theory was performed for a five-dimensional
black hole \cite{Stro}. This example provides the best laboratory
for the microscopic string theory of black holes. Besides, the
production of higher-dimensional black holes in future colliders
becomes a conceivable possibility in scenarios involving large
extra dimensions and TeV-scale gravity. Furthermore as
mathematical objects, black hole spacetimes are among the most
important Lorentzian Ricci-flat manifolds in any dimension. While
the non-rotating black hole solution to the higher-dimensional
Einstein-Maxwell gravity was found several decades ago \cite{Tan},
the counterpart of the Kerr-Newman solution in higher dimensions,
that is the charged generalization of the Myers-Perry solution
\cite{Myer} in higher dimensional Einstein-Maxwell theory, still
remains to be found analytically. Indeed, the case of charged
rotating black holes in higher dimensions has been discussed in
the framework of supergravity theories and string theory
\cite{Cvetic0,Cvetic1,Cvetic2}. Recently, charged rotating black
hole solutions in higher dimensions with a single rotation
parameter in the limit of slow rotation has been constructed in
\cite{Aliev2} (see also \cite{Aliev3,kunz1}).

On the other hand, a scalar field called the dilaton appears in
the low energy limit of string theory. The presence of the dilaton
field has important consequences on the causal structure and the
thermodynamic properties of black holes. Thus much interest has
been focused on the study of the dilaton black holes in recent
years. While exact static dilaton black hole solutions of
Einstein-Maxwell-dilaton (EMd) gravity have been constructed by
many authors ( see e.g. \cite{CDB1,CDB2,Cai,Sheykhi1}), exact
rotating dilaton black hole solutions have been obtained only for
some limited values of the dilaton coupling constant
\cite{kun,kunz2,Bri}. For general dilaton coupling constant, the
properties of  charged rotating dilaton black holes only with
infinitesimally small charge \cite{Cas} or small angular momentum
in four \cite{Hor1,Shi,Sheykhi2} and five dimensions have been
investigated \cite{Sheykhi3}. Recently, one of us has constructed
a class of charged slowly rotating dilaton black hole solutions in
arbitrary dimensions \cite{Sheykhi4}. However, in contrast to the
four-dimensional Horne and Horowitz solution, these solutions
(\cite{Sheykhi4}) have unusual asymptotics. They are neither
asymptotically flat nor (A)dS. Besides, they are ill-defined for
the string case where $\alpha=1$. The purpose of the present paper
is to generalize the four dimensional Horne and Horowitz solution
with sensible asymptotics, to arbitrary dimensions. These
asymptotically flat solutions describe an electrically charged,
slowly rotating dilaton black hole with an arbitrary value of the
dilaton coupling constant in various dimensions. We also
investigate the effects of the dilaton field on the angular
momentum and the gyromagnetic ratio of these rotating dilaton
black holes.

\section{Field equations and solutions}
We consider the $n$-dimensional $(n\geq4)$ theory in which gravity
is coupled to dilaton and Maxwell field with an action
\begin{eqnarray}
S &=&\frac{1}{16\pi }\int_{\mathcal{M}} d^{n}x\sqrt{-g}\left(
R\text{
}-\frac{4}{n-2}\partial_{\mu}\Phi \partial^{\mu}\Phi-e^{-\frac{4\alpha \Phi}{n-2}}F_{\mu \nu }F^{\mu \nu }\right)   \nonumber \\
&&-\frac{1}{8\pi }\int_{\partial \mathcal{M}}d^{n-1}x\sqrt{-h
}\Theta (h ),  \label{act1}
\end{eqnarray}
where ${R}$ is the scalar curvature, $\Phi$ is the dilaton field,
$F_{\mu \nu }=\partial _{\mu }A_{\nu }-\partial _{\nu }A_{\mu }$
is the electromagnetic field tensor, and $A_{\mu }$ is the
electromagnetic potential. $\alpha $ is an arbitrary constant
governing the strength of the coupling between the dilaton and the
Maxwell field. The last term in Eq. (\ref{act1}) is the
Gibbons-Hawking boundary term which is
chosen such that the variational principle is well-defined. The manifold $%
\mathcal{M}$ has metric $g_{\mu \nu }$ and covariant derivative
$\nabla _{\mu }$. $\Theta $ is the trace of the extrinsic
curvature $\Theta ^{ab}$ of any boundary $\partial \mathcal{M}$ of
the manifold $\mathcal{M}$, with induced metric $h _{ab}$. The
equations of motion can be obtained by varying the action
(\ref{act1}) with respect to the gravitational field $g_{\mu \nu
}$, the dilaton field $\Phi $ and the gauge field $A_{\mu }$ which
yields the following field equations
\begin{equation}
R_{\mu \nu }=\frac{4}{n-2} \partial _{\mu }\Phi
\partial _{\nu }\Phi+2e^{\frac{-4\alpha \Phi}{n-2}}\left( F_{\mu \eta }F_{\nu }^{\text{
}\eta }-\frac{1}{2(n-2)}g_{\mu \nu }F_{\lambda \eta }F^{\lambda
\eta }\right) ,  \label{FE1}
\end{equation}
\begin{equation}
\nabla ^{2}\Phi =-\frac{\alpha }{2}e^{\frac{-4\alpha
\Phi}{n-2}}F_{\lambda \eta }F^{\lambda \eta },  \label{FE2}
\end{equation}
\begin{equation}
\partial_{\mu}{\left(\sqrt{-g} e^{\frac{-4\alpha \Phi}{n-2}}F^{\mu \nu }\right)}=0. \label{FE3}
\end{equation}
We would like to find $n$-dimensional rotating solutions of the
above field equations. For small rotation, we can solve Eqs.
(\ref{FE1})-(\ref{FE3}) to first order in the angular momentum
parameter $a$. Inspection of the $n$-dimensional Kerr solutions
shows that the only term in the metric that changes to the first
order of the angular momentum parameter $a$ is $g_{t\phi}$.
Similarly, the dilaton field does not change to $O(a)$ and
$A_{\phi}$ is the only component of the vector potential that
changes. Therefore, for infinitesimal angular momentum we assume
the metric being of the following form
\begin{eqnarray}\label{metric}
ds^2 &=&-U(r)dt^2+{dr^2\over W(r)}- 2 a f(r)\sin^{2}{\theta}dt
d{\phi}\nonumber \\
 &&+ r^2 R^2(r)\left(d\theta^2 + \sin^2\theta d\phi^2+\cos^2\theta
d\Omega_{n-4}^2\right),
\end{eqnarray}
where $d\Omega^2_{n-4}$ denotes the metric of an unit $(n-4)$-
sphere. The functions $U(r)$, $W(r)$, $R(r)$ and $f(r)$ should be
determined. In the particular case $a=0$, this metric reduces to
the static and spherically symmetric cases. For small $a$, we can
expect to have solutions with $U(r)$ and $W(r)$ still  functions
of $r$ alone. The $t$ component of the Maxwell equations can be
integrated immediately to give
\begin{equation}\label{Ftr}
F_{tr}=\sqrt{\frac{U(r)}{W(r)}}\frac{Q e^{\frac{4\alpha
\Phi}{n-2}}}{\left( rR\right) ^{n-2}} ,
\end{equation}
where $Q$, an integration constant, is the electric charge of the
black hole. In general, in the presence of rotation, there is also
a vector potential in the form
\begin{equation}\label{Aphi}
 A_{\phi}=a Q C(r)\sin^2\theta.
\end{equation}
Asymptotically flat static ($a=0$) black hole solutions of the
above field equations was found in \cite{Hor2}. Here we are
looking for the asymptotically flat solutions in the case
$a\neq0$. Our strategy for obtaining the solution is the
perturbative method suggested in \cite{Hor1}.  Inserting the
metric (\ref{metric}), the Maxwell fields (\ref{Ftr}) and
(\ref{Aphi}) into the field equations (\ref{FE1})-(\ref{FE3}), one
can show that the static part of the metric leads to the following
solutions \cite{Hor2}
\begin{eqnarray}\label{U}
U(r)&=&\left[1-\left(\frac{r_{+}}{r}\right)^{n-3}\right]\left[1-\left(\frac{r_{-}}{r}\right)^{n-3}\right]^{1-\gamma\left(n-3\right)}
,
\end{eqnarray}
\begin{eqnarray}\label{W}
W(r)&=\left[1-\left(\frac{r_{+}}{r}\right)^{n-3}\right]\left[1-\left(\frac{r_{-}}{r}\right)^{n-3}\right]^{1-\gamma},
\end{eqnarray}
\begin{eqnarray}\label{R}
R(r)&=\left[1-\left(\frac{r_{-}}{r}\right)^{n-3}\right]^{\gamma/2}&,
\end{eqnarray}
\begin{eqnarray}\label{Phi}
\Phi (r)=\frac{n-2}{4}\sqrt{\gamma(2+3\gamma-n\gamma)}\ln
\left[1-\left(\frac{r_{-}}{r}\right)^{n-3}\right],
\end{eqnarray}
while the rotating part of the metric admits a solution
\begin{eqnarray}\label{f0}
f(r)&=&\left(n-3\right)\left(\frac{r_{+}}{r}\right)^{n-3}\left[1-\left(\frac{r_{-}}{r}\right)^{n-3}\right]^{\frac{n-3-\alpha^{2}}
{n-3+\alpha^{2}}}\nonumber\\
&&+\frac{(\alpha^{2}-n+1)(n-3)^{2}}{\alpha^{2}+n-3}r_{-}^{n-3}r^{2}\left[1-\left(\frac{r_{-}}{r}\right)^{n-3}\right]
^{\gamma}\nonumber\\
&& \times \int
\left[1-\left(\frac{r_{-}}{r}\right)^{n-3}\right]^{\gamma(2-n)}
\frac{dr}{r^{n}},
\end{eqnarray}
\begin{eqnarray}\label{C}
 C(r)= \frac{1}{r^{n-3}}.
\end{eqnarray}
We can also perform the integration and express the solution in
terms of hypergeometric function
\begin{eqnarray}\label{f}
f(r)&=&\left(n-3\right)\left(\frac{r_{+}}{r}\right)^{n-3}\left[1-\left(\frac{r_{-}}{r}\right)^{n-3}\right]^{\frac{n-3-\alpha^{2}}
{n-3+\alpha^{2}}}\nonumber\\
&&+\frac{(\alpha^{2}-n+1)(n-3)^{2}}{(1-n
)(\alpha^{2}+n-3)}(\frac{r_{-}}{r})^{n-3}\left[1-\left(\frac{r_{-}}{r}\right)^{n-3}\right]
^{\gamma}\nonumber\\
&& \times  _{2}F_{1} \left(\left[(n-2)\gamma,\frac
{n-1}{n-3}\right],\left[\frac {2n-4}{n-3}\right],\left({\frac
{b}{r}}\right) ^{n-3} \right).
\end{eqnarray}
Here $r_+$ and $r_{-}$ are the event horizon and Cauchy horizon of
the black hole, respectively. The constant $\gamma$ is
\begin{equation}\label{gamma}
\gamma=\frac{2\alpha^{2}}{(n-3)(n-3+\alpha^{2})}.
\end{equation}
The charge $Q$ is related to $r_+$ and $r_{-}$ by
\begin{equation}\label{Q}
Q^{2}=\frac{(n-2)(n-3)^{2}}{2(n-3+\alpha^{2})}r_{+}^{n-3}r_{-}^{n-3},
\end{equation}
and the physical mass of the black hole is obtained as follows
\cite{Fang}
\begin{equation}\label{mass}
{M}=\frac{\Omega
_{n-2}}{16\pi}\left[(n-2)r^{n-3}_{+}+\frac{n-2-p(n-4)}{p+1}r^{n-3}_{-}\right],
\end{equation}
where $\Omega _{n-2}$ denotes the area of the unit $(n-2)$-sphere
and the constant $p$ is
\begin{equation}\label{pp}
{p}=\frac{(2-n)\gamma}{(n-2)\gamma-2}.
\end{equation}
The metric corresponding to (\ref{U})-(\ref{f}) is asymptotically
flat. In the special case $n=4$, the static part of our solution
reduces to
\begin{equation}\label{U4}
U(r)= W(r)= \left( 1-{\frac {r_{+}}{r}} \right) \left( 1-{\frac
{r_{-}}{r}} \right) ^{\,{\frac {{1-\alpha}^{2}}{1+{\alpha}^{2}}}},
\end{equation}
\begin{equation}\label{R4}
R \left( r \right) = \left( 1-{\frac {r_{-}}{r}} \right) ^{{\frac
{{\alpha }^{2}}{1+{\alpha}^{2}}}},
\end{equation}
\begin{equation}\label{Phi4}
\Phi \left( r \right) =\frac{\alpha}{\left( 1+{\alpha}^{2}
\right)}\ln  \left(1- {\frac {r_{-}}{r}} \right),
\end{equation}
while the rotating part reduces to
\begin{equation}
\label{fhor} f(r)=\frac{r^{2}(1+\alpha^{2})^{2}
(1-\frac{r_{-}}{r})^{\frac{2\alpha^{2}}{1+\alpha^{2}}}}{(1-\alpha^{2})
(1-3\alpha^{2})r^{2}_{-}}-\left(1-\frac{r_{-}}{r}\right)^{\frac{1-\alpha^{2}}
{1+\alpha^{2}}}\left(1+\frac{(1+\alpha^{2})^{2}r^{2}}{(1-\alpha^{2})(1-3\alpha^{2})
r^{2}_{-}}+\frac{(1+\alpha^{2})r}{(1-\alpha^{2})r_{-}}-\frac{r_{+}}{r}\right),
\end{equation}
which is the four-dimensional charged slowly rotating dilaton
black hole solution of Horne and Horowitz \cite{Hor1}. One may
also note that in the absence of a non-trivial dilaton
($\alpha=0=\gamma $), our solutions reduce to
\begin{equation}\label{U0}
U \left( r \right) = W(r)= \left[ 1- \left( {\frac {r_{+}}{r}}
\right) ^{n-3}
 \right]  \left[ 1- \left( {\frac {r_{-}}{r}} \right) ^{n-3}
 \right],
\end{equation}
\begin{equation}\label{f0}
f \left( r \right)
=(n-3)\left[\frac{r^{n-3}_{-}+r^{n-3}_{+}}{r^{n-3}}-\left(\frac{r_{+}r_{-}}{r^{2}}\right)^{n-3}\right],
\end{equation}
which describe $n$-dimensional Kerr-Newman black hole in the limit
of slow rotation \cite{Aliev2}.

Next, we calculate the angular momentum and the gyromagnetic ratio
of these rotating dilaton black holes which appear in the limit of
slow rotation parameter. The angular momentum of the dilaton black
hole can be calculated through the use of the quasilocal formalism
of the Brown and York \cite{BY}. According to the quasilocal
formalism, the quantities can be constructed from the information
that exists on the boundary of a gravitating system alone. Such
quasilocal quantities will represent information about the
spacetime contained within the system boundary, just like the
Gauss's law. In our case the finite stress-energy tensor can be
written as
\begin{equation}
T^{ab}=\frac{1}{8\pi }\left(\Theta^{ab}-\Theta h ^{ab}\right) ,
\label{Stres}
\end{equation}
which is obtained by variation of the action (\ref{act1}) with
respect to the boundary metric $h _{ab}$. To compute the
angular momentum of the spacetime, one should choose a spacelike surface $%
\mathcal{B}$ in $\partial \mathcal{M}$ with metric $\sigma _{ij}$,
and write the boundary metric in ADM form
\[
\gamma _{ab}dx^{a}dx^{a}=-N^{2}dt^{2}+\sigma _{ij}\left( d\varphi
^{i}+V^{i}dt\right) \left( d\varphi ^{j}+V^{j}dt\right) ,
\]
where the coordinates $\varphi ^{i}$ are the angular variables
parameterizing the hypersurface of constant $r$ around the origin,
and $N$ and $V^{i}$ are the lapse and shift functions
respectively. When there is a Killing vector field $\mathcal{\xi
}$ on the boundary, then the quasilocal conserved quantities
associated with the stress tensors of Eq. (\ref{Stres}) can be
written as
\begin{equation}
Q(\mathcal{\xi )}=\int_{\mathcal{B}}d^{n-2}\varphi \sqrt{\sigma }T_{ab}n^{a}%
\mathcal{\xi }^{b},  \label{charge}
\end{equation}
where $\sigma $ is the determinant of the metric $\sigma _{ij}$, $\mathcal{%
\xi }$ and $n^{a}$ are, respectively, the Killing vector field and
the unit normal vector on the boundary $\mathcal{B}$. For
boundaries with rotational ($\varsigma =\partial /\partial \varphi
$) Killing vector field, we can write the corresponding quasilocal
angular momentum as follows
\begin{eqnarray}
J &=&\int_{\mathcal{B}}d^{n-2}\varphi \sqrt{\sigma
}T_{ab}n^{a}\varsigma ^{b},  \label{Angtot}
\end{eqnarray}
provided the surface $\mathcal{B}$ contains the orbits of
$\varsigma $. Finally, the angular momentum of the black holes can
be calculated by using Eq. (\ref{Angtot}). We find
\begin{equation}
{J}=\frac{a\Omega
_{n-2}}{8\pi}\left(r^{n-3}_{+}+\frac{(n-3)(n-1-\alpha^{2})r^{n-3}_{-}}{(n-3+\alpha^{2})(n-1)}\right).
\label{J}
\end{equation}
For $a=0$, the angular momentum  vanishes, and therefore $a$ is
the rotational parameter of the dilaton black hole. In the case
$n=4$, the  angular momentum reduces to
\begin{equation}
{J}=\frac{a}{2}\left(r_{+}+\frac{3-\alpha^2}{3(1+\alpha^2)}r_{-}\right),
\label{J4}
\end{equation}
which restores the angular momentum of the four-dimensional Horne
and Horowitz solution \cite{Hor1}, while in the absence of dilaton
field $(\alpha=0)$, the angular momentum reduces to
\begin{equation}
{J}=\frac{a\Omega _{n-2}}{8\pi}\left(r^{n-3}_{+}+r^{n-3}\right),
\label{Jn}
\end{equation}
which is the angular momentum  of the $n$-dimensional Kerr-Newman
black hole. Next, we calculate the gyromagnetic ratio of this
rotating dilaton black holes. The magnetic dipole moment for this
asymptotically flat slowly rotating dilaton black hole can be
defined as
\begin{equation}\label{mu}
{\mu}=Qa.
\end{equation}
The gyromagnetic ratio is defined as a constant of proportionality
in the equation for the magnetic dipole moment
\begin{equation}\label{mu}
{\mu}=g\frac{QJ}{2M}.
\end{equation}
 Substituting $M$ and $J$ from Eqs. (\ref{mass}) and  (\ref{J}),
 the gyromagnetic ratio $g$ can be obtained as
\begin{equation}\label{g}
g=\frac{(n-1)(n-2)[(n-3+\alpha^2)r^{n-3}_{+}+(n-3-\alpha^2)r^{n-3}_{-}]}{
(n-1)(n-3+\alpha^2)r^{n-3}_{+}+(n-3)(n-1-\alpha^2)r^{n-3}_{-}}.
\end{equation}
\begin{figure}[tbp]
\epsfxsize=7cm \centerline{\epsffile{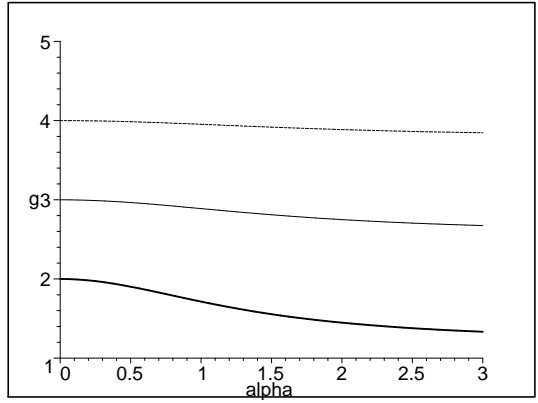}} \caption{The
behaviour of the gyromagnetic ratio $g$ versus $\protect\alpha$ in
various dimensions for $r_{-}=1$, $r_{+}=2$. $n=4$ (bold line),
$n=5$ (continuous line), and $n=6$ (dashed line).} \label{figure1}
\end{figure}
It was argued in \cite{Hor1} that the dilaton field modifies the
gyromagnetic ratio of the asymptotically flat four dimensional
black holes. Our general result here in $n$-dimensions confirms
their arguments. We have shown the behaviour of the gyromagnetic
ratio $g$ of the dilatonic black holes versus $\protect\alpha$ in
figure \ref{figure1}. From this figure we find out that the
gyromagnetic ratio decreases with increasing $\alpha$ in any
dimension. In the absence of a non-trivial dilaton
$(\alpha=0=\gamma)$, the gyromagnetic ratio reduces to
\begin{equation}\label{gkerr-newman}
{g}=n-2,
\end{equation}
which is the gyromagnetic ratio of the $n$-dimensional Kerr-Newman
black hole (see e.g.\cite{Aliev2}). When $n=4$, Eq. (\ref{g})
reduces to
\begin{equation}\label{gHor}
{g}=2-\frac{4\alpha^{2}r_{-}}{(3-\alpha^{2})r_{-}+3(1+\alpha^{2})r_{+}},
\end{equation}
which is the gyromagnetic ratio of the four dimensional Horne and
Horowitz dilaton black hole.

\section{Summary and Conclusion}
To sum up, we found a class of asymptotically flat slowly rotating
charged dilaton black hole solutions in higher dimensions. Our
strategy for obtaining this solution was the pertarbative method
suggested by Horne and Horowitz \cite{Hor1} and solving the
equations of motion up to the linear order of the angular momentum
parameter. We stared from the asymptotically flat non-rotating
charged dilaton black hole solutions in $n$-dimensions
\cite{Hor2}. Then, we considered the effect of adding a small
amount of rotation parameter $a$ to the solution. We discarded any
terms involving $a^2$ or higher powers in $a$. Inspection of the
Kerr-Newman solutions shows that the only term in the metric which
changes to $O(a)$ is $g_{t\phi}$. Similarly, the dilaton field
does not change to $O(a)$. The vector potential is chosen to have
a non-radial component $A_{\phi} = aQC(r)\sin^2{\theta} $ to
represent the magnetic field due to the rotation of the black
hole. As expected, our solution $f(r)$ reduces to the Horne and
Horowitz solution for $n=4$, while in the absence of dilaton field
$(\alpha=0=\gamma)$, it reduces to the $n$-dimensional Kerr-Newman
modification thereof for small rotation parameter \cite{Aliev2}.
We calculated the angular momentum $J$ and the gyromagnetic ratio
$g$ which appear up to the linear order of the angular momentum
parameter $a$. Interestingly enough, we found that the dilaton
field modifies the value of the gyromagnetic ratio $g$ through the
coupling parameter $\alpha$ which measures the strength of the
dilaton-electromagnetic coupling. This is in agrement with the
arguments in \cite{Hor1}.

\acknowledgments{This work has been supported financially by
Research Institute for Astronomy and Astrophysics of Maragha,
Iran.}


\end{document}